\documentclass[journal]{IEEEtran}
\usepackage{graphicx}
\usepackage{tabularx}
\usepackage[justification=centering]{caption}
\usepackage{cite}
\usepackage{hyperref} 
\usepackage{latexsym}
\usepackage{amsmath}
\usepackage{algorithm}
\usepackage[noend]{algpseudocode}
\usepackage{booktabs}
\usepackage{multirow}
\usepackage{siunitx}
\usepackage{pdfpages}
\usepackage{booktabs}
\usepackage{afterpage}

\usepackage{float}
\usepackage{subfigure}

\makeatletter
\def\BState{\State\hskip-\ALG@thistlm}
\makeatother

\renewcommand{\baselinestretch}{0.90}

\ifCLASSINFOpdf
\else
\fi

\begin{document}
%
% paper title
% can use linebreaks \\ within to get better formatting as desired
\title{A Novel Key Management and Data Encryption Method for Metering Infrastructure of Smart Grid}
%
%
% author names and IEEE memberships
% note positions of commas and nonbreaking spaces ( ~ ) LaTeX will not break
% a structure at a ~ so this keeps an author's name from being broken across
% two lines.
% use \thanks{} to gain access to the first footnote area
% a separate \thanks must be used for each paragraph as LaTeX2e's \thanks
% was not built to handle multiple paragraphs
%

\author{Imtiaz Parvez,~\IEEEmembership{Student Member,~IEEE,}
        Arif I. Sarwat,~\IEEEmembership{Member,~IEEE,}
        My T. Thai,~\IEEEmembership{Senior Member,~IEEE,}
         ~and~Anurag K. Srivastava,~\IEEEmembership{Senior Member,~IEEE}% <-this % stops a space
%\thanks{Imtiaz Parvez and Arif I. Sarwat  are with the Department
%of Electrical and Computer Engineering, Florida International University, FL,
%33174 USA e-mail:iparv001@fiu.edu, asarwat@fiu.edu}% <-this % stops a space
%\thanks{Ganesh Kumar Venayagamoorthy is with the Department
%of Electrical and Computer Engineering, Clemson University, SC,
% 29634-0915 USA e-mail:gvenaya@clemson.edu.}}% <-this % stops a space
%\thanks{Manuscript received April 19, 2005; revised January 11, 2007.}
\thanks{This research was supported in part through U.S. National Science Foundation under the grant RIPS-1441223.}
}

\maketitle

\begin{abstract}
%\boldmath
In the smart  grid, smart meters, and numerous control and monitoring applications employ bidirectional  wireless  communication, where  security is  a  critical   issue.  In  key  management  based  encryption method  for the smart  grid,  the Trusted  Third  Party  (TTP),  and  links between  the smart  meter  and  the third  party  are  assumed  to be fully trusted and reliable.  However, in wired/wireless  medium, a man-in-middle may want  to interfere, monitor  and  control  the  network, thus  exposing its  vulnerability.  Acknowledging  this, in this paper, we  propose  a  novel  two level encryption   method  based  on  two  partially   trusted  simple servers (constitutes the TTP) which implement this method without increasing  packet  overhead.  One server is responsible for data encryption between the meter and control center/central database, and the other server manages  the  random sequence of data  transmission. Numerical calculation shows that the number of iterations required to decode a message is large which is quite impractical. Furthermore, we introduce One-class  support  vector  machine  (machine learning) algorithm  for node-to-node authentication utilizing the location information and the data  transmission history (node identity, packet size and frequency of transmission). This secures data communication  privacy  without  increasing the complexity  of the conventional  key management scheme.
\end{abstract}

% IEEEtran.cls defaults to using nonbold math in the Abstract.
% This preserves the distinction between vectors and scalars. However,
% if the journal you are submitting to favors bold math in the abstract,
% then you can use LaTeX's standard command \boldmath at the very start
% of the abstract to achieve this. Many IEEE journals frown on math
% in the abstract anyway.

% Note that keywords are not normally used for peerreview papers.


% For peer review papers, you can put extra information on the cover
% page as needed:
% \ifCLASSOPTIONpeerreview
% \begin{center} \bfseries EDICS Category: 3-BBND \end{center}
% \fi
%
% For peerreview papers, this IEEEtran command inserts a page break and
% creates the second title. It will be ignored for other modes.
\IEEEpeerreviewmaketitle

\section{Introduction}

Smart grid is the modern power system infrastructure whose distribution system is  upgraded  via bidirectional communication, and pervasive and intelligent computing capabilities in order to provide improved control, efficiency, reliability and safety. In the arena of smart grid, the Advanced Metering Infrastructure (AMI) is the distribution level building block. It consists of millions of meters interlinked by a hierarchical or mesh-connected or hybrid network. The meters communicate with the control center in a wireless fashion. The typical communication standards for the AMI are ZigBee, WiFi and LTE \cite{j42}. In the AMI, smart meters collect and store the instantaneous energy consumption and  report  it  periodically to  the back office (the  control center) as against reporting of monthly total energy consumption recorded in the conventional meters. It also allows the consumers to actively engage in electricity trade by selling home- generated, unused electricity (from solar panel, windmill etc.) to the grid. AMI also outfits the service provider with control and monitoring equipment, including outage management, demand response, and disaster prevention and recovery information.

Since AMI has to use wireless communication, security issues become a critical problem. Observing the  usage patterns, an adversary/thief can predict the presence of the targeted consumers at home  which  can  be  a  threat  to civil lives and privacy. Furthermore, from the fine grained energy consumption data, the home appliance companies receive  information about the life style patterns of consumers and the energy utilization of their home appliances. Thus  competing companies  can use this valuable information  in their businesses. The consumers might want to alter the consumption data to reduce their electricity bill. The most crucial thing is that the opponent/hacker might jam or take over the AMI network by sending false signal to meters on an unsecured system, which may cause power outage in a wide area as well as an imbalance in the demand generation model.

Like  all  other systems, AMI  needs  to  comply with the  requirements  of  security- confidentiality, integrity, availability and accountability (non-repudiation) \cite{j1,j17}. Confidentially implies the accessibility  of  data  by  any authorized entity, and any intentional or unintentional disclosures of data must be denied. Since the main objective of the AMI is to collect energy consumption information that conveys consumer life patterns, habits and energy usage, this data must be concealed. Integrity requires reflecting authentic data correctly without any modifications, additions, or deletions. If any unauthorized subject performs (or attempts to perform) such actions, this must be  detected. Since not only the hacker, but also the consumer, might want to alter the consumption data by trapping and resending, the integrity of data is very important. Availability requires the accessibility of data by an authorized user on demand. If  the required data  is  not  found  at  the  time  of need, the system violates the availability aspect of the security requirement of the system. Any natural or intentional incidents (i.e. hacking) must not hamper the system from operating correctly. For example, if the hacker wants to jam the network, the system must comply with the availability aspect. Accountability (non-repudiation) means non-deniability of an action, i.e., the entities used in receiving or transmitting data must not deny it. If an entity does not receive data, it will not, subsequently, state that it has received it. In the AMI network, accountability ensures a timely response to the command and control, and integrity of billing profile, etc.

% Along with data-centric security, the AMI network needs to comply with the behavioral approach. The behavioral approach attributes include safety, correctness, performance, reliability, trustworthiness, and dependability. In this paper, we focus on  information centric security and privacy.

\setlength\belowcaptionskip{-4ex}

%\begin{figure*}
%\includegraphics[width=\textwidth,height=7cm]{AMIdiagrampdf.pdf}
%  %\includegraphics[width=\textwidth]{AMIdiagrampdf.pdf}
%  \caption{AMI architecture.}
%\end{figure*}

The main drawback of implementing security scheme in AMI is the limited memory and low computational ability of the smart meters. Furthermore, AMI is a huge network, consists of millions of meters. So we need a lightweight but robust security scheme. In the literature, the key management-based encryption method has been proposed as a prominent security scheme for smart grid, which includes a Trusted Third Party (TTP) \cite{j8,j19,j21}. In all TTP management systems, it is assumed that the TTP is fully trusted. However, the TTP itself, and the meters and communication links among the TTP and the meters could also be compromised.  

Based on the semi-trusted servers of the TTP and untrustworthy/unreliable communication links, in this paper, we propose a key management scheme which includes a node (meter) to the Service Provider (SP) data encryption as well as randomized packet transmission. In our early work \cite {j41}, the scheme consists of two independent servers. The master server manages the public-private key before data is sent to the back office of the SP. Additionally, the auxiliary server receives the random sequence of data from the smart meter, in response to the public key sent by the master server. Both the private key associated with public key and the random sequence are used to retrieve the data at the back office. In this paper, we extend \cite {j41} by using Received Signal Strength (RSS) and One class Support Vector Machine (OCSVM) techniques  for node to node authentication. RSS is used for localization of meters using the received signal strength from neighbor meters. On the other hand, OCSVM is used to detect new and outlier data/packet, using current and previous data transmission history, which has not been considered in the literature to our best knowledge. The introduction of two separate servers for key management and random sequenced packet transmission increases robustness in security  in untrustworthy communication medium and servers. Additionally, OCSVM based node authentication increases sturdiness in key management system without increasing any overhead and it is easy to implement in devices with limited memory and computational ability.

%Building over this foundation, here we present an extended work. For node-to-node authentication, we use random signal strength (RSS) and machine learning (One class Support Vector Machine (One-class  SVM)) based  authentication  method. One-class SVM detects novelty and outliers using current and previous data transmission history. To the best of our knowledge, this unique encryption method has never been proposed.
%The rest of the contents are organized as follows. In section II, security threats for the AMI are  described in  detail. Section III describes the literature review of the security schemes which are proposed for the AMI. In  section IV,  details of  the proposed structure of the AMI are illustrated. In section V, the theoretical background of RSS based localization and One-class Support vector machine (One- class SVM) are presented. Section VI presents the communication and traffic flow among smart meters, servers and service providers. Furthermore, in section VI, simulation results and theoretical security strength are analyzed. Finally, a brief conclusion is included in section VII.

The rest of the paper is organized as follows. Section~II describes the literature review of the security schemes proposed for the AMI. In  Section~III,  details of  the proposed structure of the AMI are introduced. In Section~IV, the theoretical background of RSS based localization, OCSVM and entropy of data packet are presented. Section~V presents the communication and traffic flow among smart meters, servers and SPs. Furthermore, in Section~VI, simulation results and theoretical security strength of a data packet are analyzed. Finally, concluding remarks are provided in Section~VII.

\section{Literature review}

In  the recent years, security issues in AMI attracted significant attention of different communities (electrical engineers, computer science graduates, IT experts etc.) due to extensive use of wireless communication. In \cite{j12}, a 128-bit Asymmetric Encryption Scheme Galois Counter Mode (AES GCM) cryptography based security IC has been proposed and performance comparison between the hardware-based and software-based crypto-engines is presented. In \cite {j4,j5}, a game theory based security system has been introduced where two parties (an attacker and a defender) interact pro-actively and reactively for different attack severity levels. In \cite{j34}, several nodes are selected randomly for intermediate node-to-node authentication. Even though this reduces the packet overhead, it still has vulnerability in the middle of communication. In \cite{j55}, randomization of the AMI configuration has been proposed to make its behavior unpredictable to the hacker, whereas the behavior is predictable to the control center. In \cite{j13}, Costas $\textit{et al.}$ introduced anonymization of data by randomizing node identity using a TTP. But communication overhead is increased due to the need for the TTP  to communicate with all nodes simultaneously. Moreover, retrieving information from a large network becomes complicated. In \cite{j14}, homomorphic encryption has been introduced, which also has the problem of retrieving data from a large network being complicated. 

In\cite{j32}, an Identity Based Signcryption (IBS) for zero configuration encryption and authentication has been proposed for end-to-end communication solution. In\cite{j33}, a node-to-node encryption with its own secret key has been proposed. But for a large network, the packet overhead increases for both IBS and node-to-node authentication. To distribute the keys and manage the network, a wireless sensor network based Public Key management Infrastructure (PKI) has been proposed in \cite{j7} which has problem of generation large number of unique keys. In \cite{j15}, a key management system has been introduced based on DLMS/COSEM standard providing two main information security features: data access security and data transport security. Since DLMS/COSEM is an open standard and allows a number of variations in the protocol implementation, it increases the complexity in the client side. In\cite{j16}, a two layer security scheme for meter to Data Concentrator (DC) and DC to control center of SP has been provided in Taiwan. For meter to DC, IEC 62056 based encryption method, and for DC to the back office, public key management system has been proposed. But here in each time step, encryption and decryption need to be performed twice. 

Different than these earlier work in the literature, in our proposed scheme, we provide  improvements in the key management schemes for AMI systems. In particular, in our approach, we introduce randomization of data transmission  as well as node-to-node authentication method considering unreliable communication scenario and using machine learning technique. The introduction of randomization and the machine learning based node authentication increases the robustness of the key management system without increasing any overhead.

\section{Architecture of AMI}

\begin{figure}[t]
	\centering
	\includegraphics[width=1\linewidth]{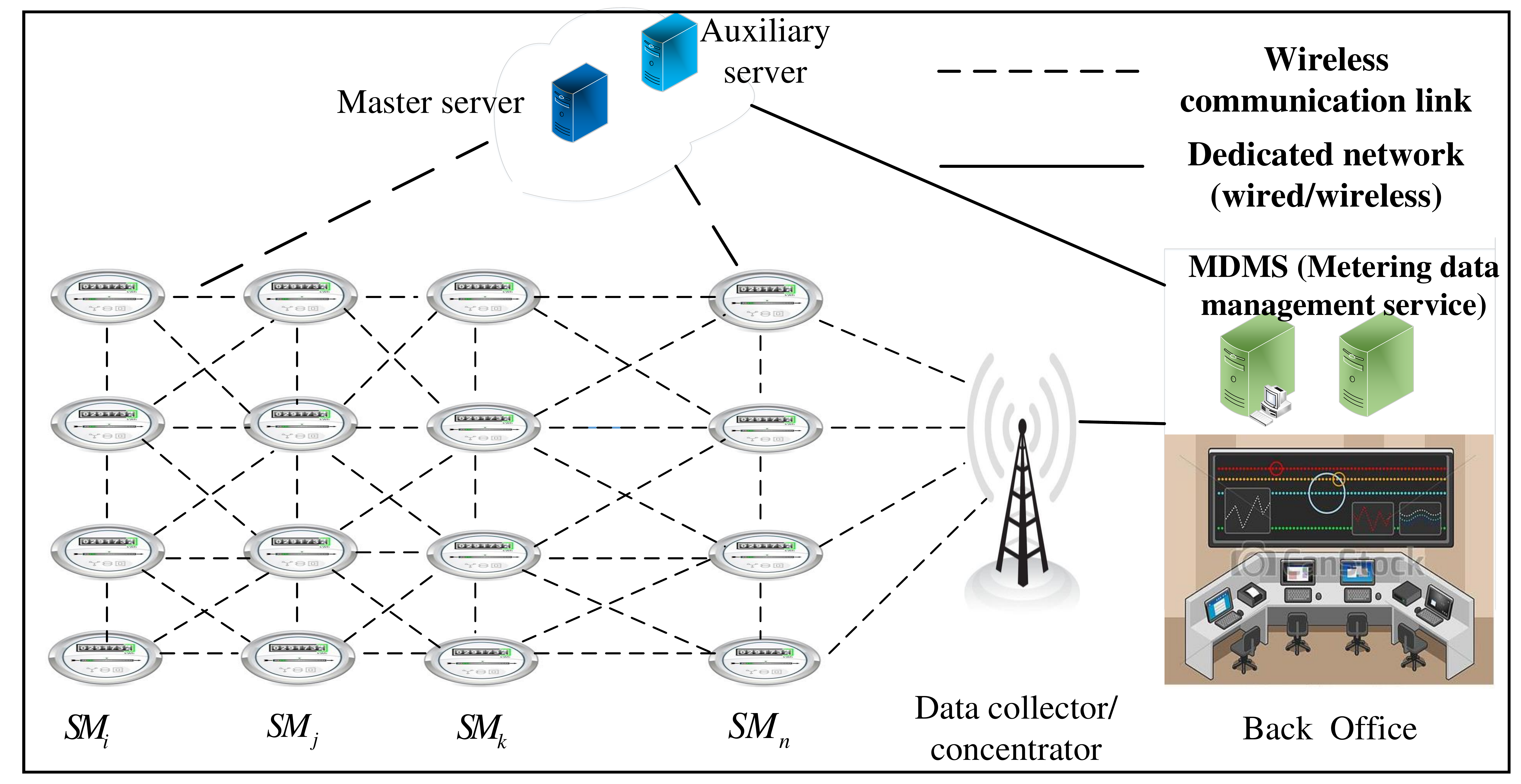}
	\caption{AMI architecture consisting of a cluster of mesh/hybrid connected meters, data concentrator, KMS and control center.}
	\label{fig:Packet}
\end{figure}

The AMI is a web like network with millions of meters as shown in Fig. 1. In here we provide the insights into AMI architecture, which encompasses components including home appliances, control center of utility service provider and security infrastructure.

%The responsibility of AMI includes collection of consumption data as well as management of power distribution. It comprises of different components with various applications.

\begin{itemize}
	
	\item \textit{Smart meter:} It is a solid state device responsible for collecting, storing and sending data to the back office using wireless communication and at interval less than 1 hour. The home appliances are connected to a smart meter by a network, forming Home Area Network (HAN).
	
	\item \textit{Neighborhood Area Network (NAN):} The meters are connected among themselves through a mesh or hierarchical or hybrid  network termed as NAN. The network may be wired (e.g. PLC) or  wireless (e.g. WiFi, ZigBee, GPRS). The head end of the NAN is the data concentrator or gateway which is connected to back office by a dedicated wired or wireless connection (optical fiber, cellular network, etc.).
	
	\item \textit{Control center/Hardware and software control system/Utility back office/SP:} The control center receives the consumption data from the AMI network and uses them for making bills. This fine grained data can be used to optimize power generation and distribution. Besides that, electricity distribution, control, and monitoring is performed from the control center.
	
	\item \textit{Master server:} In our architecture, the master server is semi- trusted. When a meter requests the key, the master server generates a public key and private key pair. The master server uni-casts the public key to the specific meter for data encryption and the corresponding private key is sent to the auxiliary server and  control center for data decryption.
	
	\item \textit{Auxiliary server:} Before the encryption of data with the public key sent by the master server, the smart meter generates a random sequence. This random sequence is sent as encrypted by public key to auxiliary server. The auxiliary server receives random sequence and authenticates it by the private key of smart meter before forwarding it to the control center for final decryption.

\end{itemize}

\section{OC-SVM and RSS Algorithm, and entropy of a data packet}

In our scheme, we use RSS for the localization of meter, and OC-SVM for node-to-node authentication. In the rest of this section, we describe the two algorithms, and security strength calculation of a data packet by entropy in further detail.

%In this section,  localization using random signal strength (RSS)\cite{j23,j24,j25} and classification using LDA \cite{j26,j27} has been explained.

\subsection{Received Signal Strength (RSS) based localization}

RSS based localization can be used in localizing a new meter in collaboration with meters whose positions are already known. The location information is used to authenticate the source node using OCSVM.
 
Let us consider an unknown positioned meter at a location $(x,y)$ accompanied by partially dispersed known position meters at locations $(x_l,y_l)$, where $1\leq{l}\leq{n}$. The received signal strength at location $(x_l,y_l)$ can be denoted by $\psi_l$
\begin{equation}
\psi_l=c-10\gamma\log(d_l)+w_l,
\end{equation}
where $c$ is an unknown constant that depends on transmitted power, frequency etc, and $\gamma$ is the path loss constant. For a lossy environment, the typical value of $\gamma$ is 4-6. In our model, $\gamma=2.93$ has been used considering residential area. The parameter $d_l$ is the euclidean distance between the known and unknown position meter defined as follows:
\begin{equation}
d_l= \sqrt{(x-x_l)^2 + (y-y_l)^2},
\end{equation}
and $w_l$ is the zero mean random Gaussian noise with standard deviation $\sigma_l$. The value of $\sigma_l$ ranges from 6 to 12 dB.\\

Let us define, the $\boldsymbol{\theta}$ and $\boldsymbol{\psi}$ as
$\boldsymbol{\theta}=[x,y,z]^T$ and $\boldsymbol{\psi}=[\psi_1,\psi_2,.......,
\psi_n]^T$.\\

\noindent The likelihood function of  $\boldsymbol{\theta}$  for a given RSS measurement $\boldsymbol{\psi}$, $f(\boldsymbol{\theta|\psi)}$ is given by
\begin{equation}
f(\boldsymbol{\theta|\psi)}=c_1 \exp \Bigg\{-\sum_{l=1}^{n} \frac{\{\psi_l -c+10\gamma\log(d_l)\}^2}{2\sigma_l^2}\Bigg\},
\end{equation}
where $c_1$ is a constant.

\noindent The Maximum Likelihood (ML) estimate of $\boldsymbol{\theta}$, denoted by $\boldsymbol{\hat{\theta}}$, can be found from the following equation
\begin{equation}
\begin{split}
\hat{\boldsymbol{\theta}}&= \arg\max f(\boldsymbol{\theta|\psi)} \\
           &=\arg\min\Bigg\{-\sum_{l=1}^{n} \frac{\{\psi_l -c+10\gamma\log(d_l)\}^2}{2{\sigma_l}^2}\Bigg\} ,
\end{split}
\end{equation}
The above equation is an optimization problem. Various optimization techniques such as differential evolution, dynamic relaxation and particle swarp optimization (PSO)  can be used to solved (4). In our problem, we used PSO to solve the non-linear optimization problem. Finally, the ML estimator yields  the location $(x,y)$ and reference power $z$ of the unknown positioned meter
\begin{align}
(x,y,z)=\{\hat{\theta} (1), \hat{\theta}(2), \hat{\theta}(3)\}.
\end{align}

\subsection{OCSVM algorithm}

The Support Vector Machine (SVM) is a machine learning algorithm based on modern statistical learning theory \cite{j36,j37}. It separates two classes by constructing a hyper surface in the input space. In this input space, the input is mapped to a higher dimensional feature space by non-linear mapping. In this section, we describe OCSVM for segregating outliers  from a cluster of data.  

Let us consider a data space  $\boldsymbol{\Psi}={(x_i,y_i)}\in\mathcal{R}^d$ $i=\{1,2,3,....,n\}$ where $x_i\in \mathcal{R}^d$ is the input data and $y_i\in \{-1,+1\}$  is the corresponding output pattern in the dedicating class membership. SVM first projects the input vector $x$ to a higher dimensional space $\mathcal{H}$ by a non-linear operator $\Phi(\cdot): R^n\longleftrightarrow\mathcal{H}$ where the data projection is linearly separable. The non-linear SVM classification is defined as 

\begin{equation}
\Omega(x)=w^T \Phi(x) +b,    w\in \mathcal{H} , b\in \mathcal{R},
\end{equation}
which is linear in terms of projected data $\Phi(x)$ and non-linear in terms of original data $x$.

\begin{figure}[h!]
\centering
\includegraphics[width=0.65\linewidth]{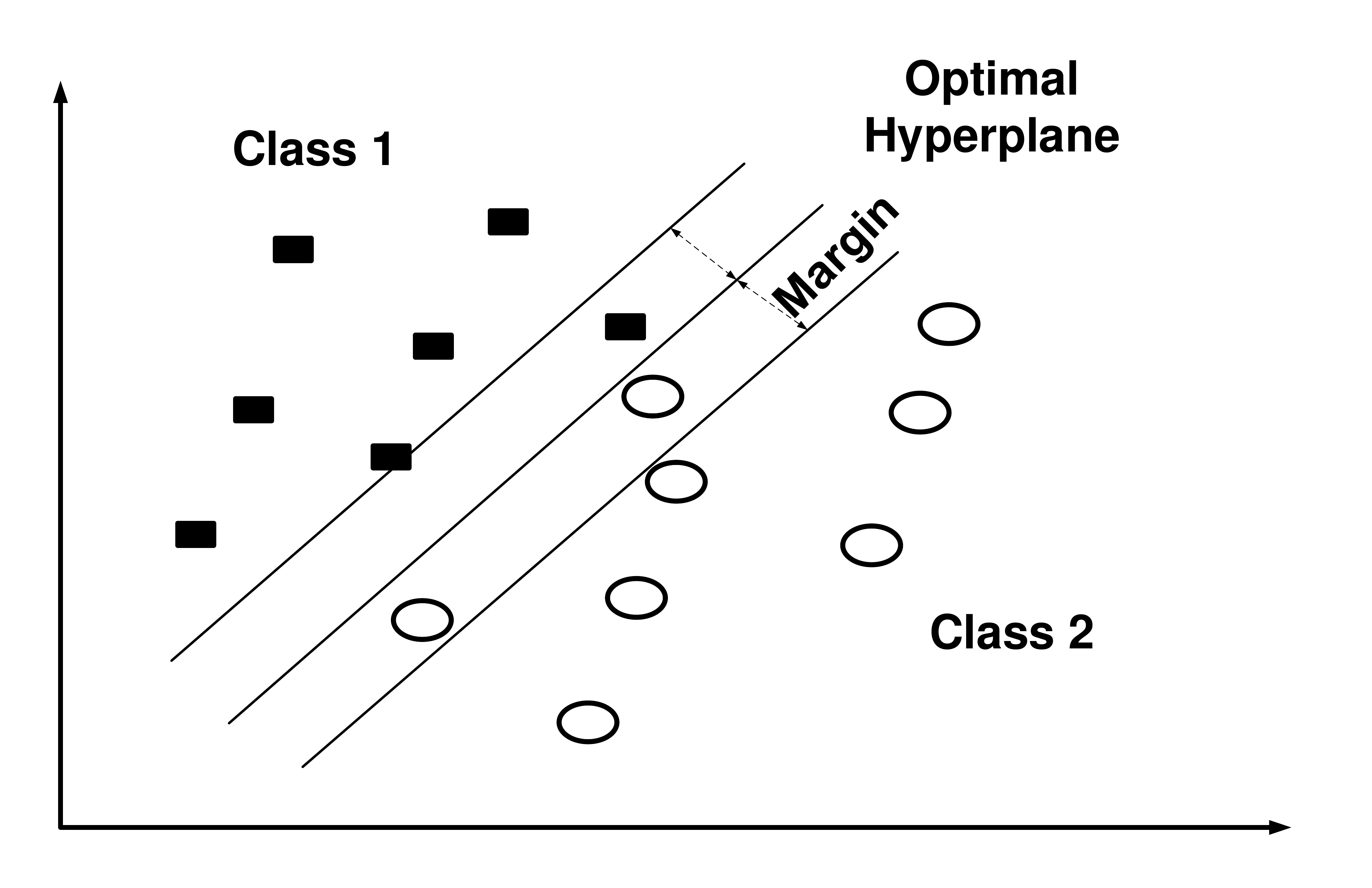}
\caption{Separation of two class in hyperplane.}
\label{fig: Packet}
\end{figure}

\vspace{1\baselineskip}

SVM tries to maximize the separation margin between two classes of data in hyperplane as shown in Fig. 2. To prevent SVM from over fitting with noisy data (or create soft margin), slack variables ($\xi_i$) are introduced to allow some data to lie within the margin. Then the objective function which includes the minimization of $\lVert w \rVert$ can be written as 
\begin{equation}
\min_{w,b,\xi_i} \frac{\lVert w \rVert}{2} + c \sum_{i=1}^n \xi_i,
\end{equation}

%\begin{equation}
%\begin{split}
%y_i(w^T\phi(x_i) +b)\geq 1- \xi_i \\
%\xi_i\geq 0
%\end{split}
%\end{equation}

\begin{equation}
\begin{array}{l}
{\rm such~that~}
\displaystyle y_i(w^T\Phi(x_i) +b)\geq 1- \xi_i, \\
{\rm and~}
\displaystyle \xi_i\geq 0 .\\
\end{array} 
\label{eq: xdef1}
\end{equation}
where $c>0$  is the regularization parameter that determines the trade-off of maximizing the margin and the number of training data within the margin (thus reducing the training errors). To minimize the objective function of (7) by using the Lagrange multipliers technique, the necessary condition for $w$ is
 
\begin{equation}
w=\sum_{i=1}^{n} \gamma_i y_i \Phi_i(x)
\end{equation}
where $\gamma_i>0, i=0,1,2.....n$ are Lagrange multipliers corresponding to the constraints in equation (8).
 
The Lagrange multipliers ($\gamma_i$) can be solved from equation (7) and written as 

\begin{equation}
\max w(\gamma_i) = \sum_{i=1}{n} \gamma_i -\frac{1}{2} \sum_{i=1}{n}\sum_{j=1}{n} \gamma_i \gamma_j y_i y_j k(x_i,x_j)
\end{equation}

\begin{equation}
\begin{array}{l}
{\rm such~that~}
\displaystyle 0\leq \gamma_i\leq c, \\
{\rm and~}
\displaystyle \sum_{i=1}^{n} \gamma_i y_i=0 .\\
\end{array} 
\label{eq: xdef}
\end{equation}

In (10), $k(x,y)=\Phi(x)^T \Phi(y)$ is known as the kernel function. It determines the mapping of the input vector to a high dimensional feature space. In this paper, we consider two kernel functions. The Guassian RBP kernel is given by: \\

$\textrm {Gaussian RBF kernel:} \hspace{0.1cm} k(x,y)=\exp (\frac{{\lVert x-y \rVert}^2 }{\sigma})$\\

\noindent while the Polynomial kernel is given by: \\

$\textrm {Polynomial kernel:} \hspace{0.1cm}  k(x,y)=((xy)+1)^p$

\noindent where $p\in \mathcal{N}$ is the degree of the polynomial function and
$\sigma\in\mathcal{R}$ is the width of the RBF function.\\

OCSVM detects abnormal data within a class\cite{j37}. OCSVM  maps the input vector to feature dimension according to the kernel function and separates it from the origin with maximum margin. It penalizes the outliers by employing slack variables $\xi$ in the objective function and controls carefully the trade off between empirical risk and regularize the penalty.

\begin{figure}[h]
\centering
\includegraphics[width= 0.65\linewidth]{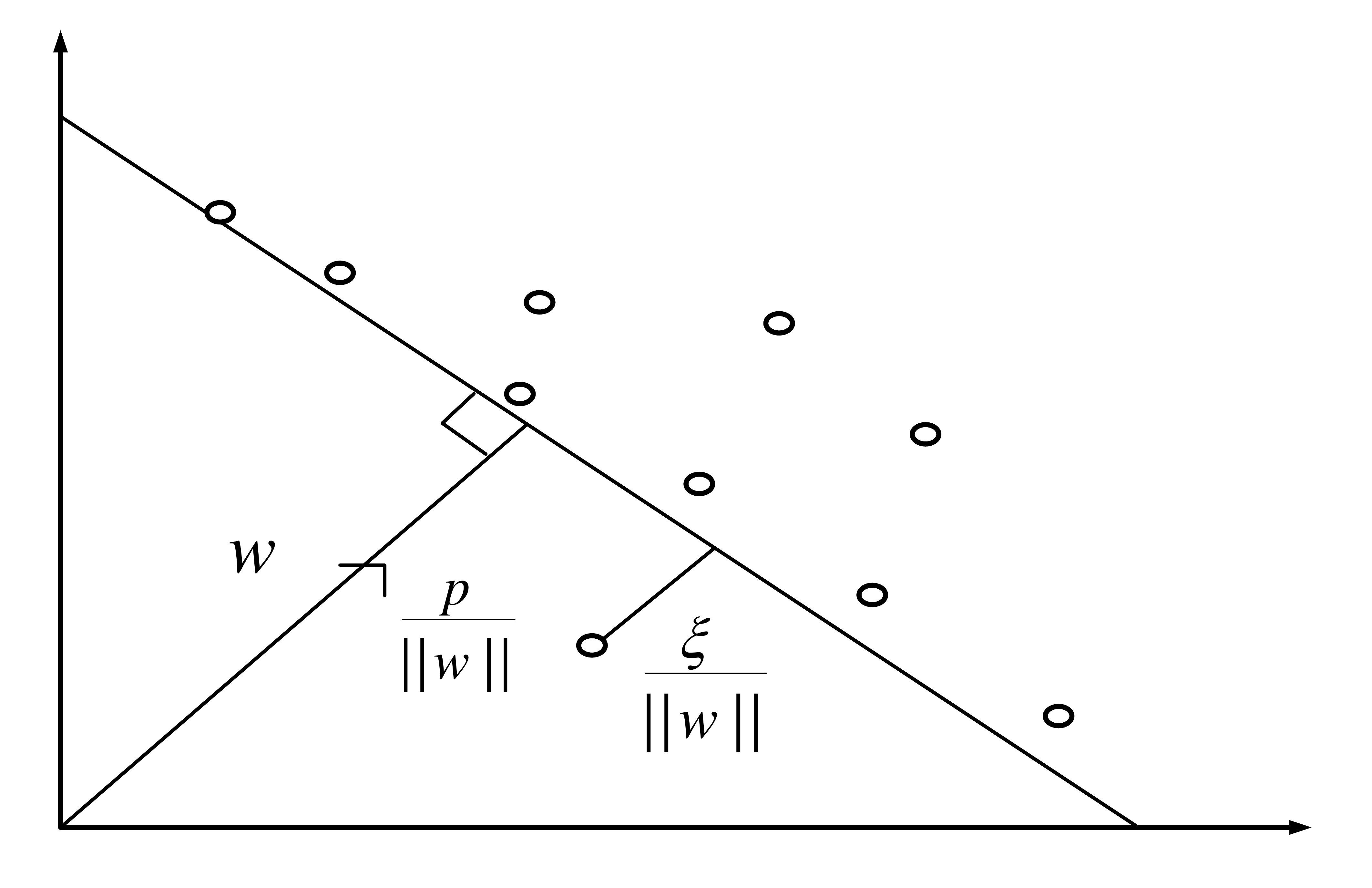}
\caption{Feature dimensional space for OCSVM }
\label{fig: Packet}
\end{figure}

The quadratic programming minimization function

\begin{equation}
\min_{w,\xi_i,\rho} \frac{1}{2} {\lVert w\rVert}^{2} + \frac{1}{vn} \sum_{i=1}{n} \xi_i-\rho
\end{equation}

\begin{equation}
\begin{array}{l}
\textrm{such that~}
\displaystyle (w\cdot \Phi (x_i) \geq \rho -\xi_i,\\
\textrm{and~}
\displaystyle \xi_i \geq 0 ,\hspace{2cm} i=1,2,3,....,n.\\
\end{array} 
\label{eq:xdef}
\end{equation}
where $\Phi$ is the kernel function for mapping, $\xi_i$ is the slack variables,
$v\in (0,1]$ is a prior fixed constant, and $\rho$ is the decision value that determines whether a given point falls within the estimated high density region.
Then the resultant  decision function $f_{w,p}^m (x)$  takes the form 

\begin{equation}
f(x)=\text{sgn} (w^{*T}\Phi(x)-\rho^{*})
\end{equation}
where $\rho^{*}$ and $w^{*}$ are the values of $w$ and $\rho$ solving from the equation (12).

In OCSVM, $v$ characterizes the solution instead of $c$ (smoothness operation)-

\begin{itemize}

	\item It determines an upper bound on the fraction of outliers.
	
		\item It is the lower bound on the number of training examples used as support vector.
\end{itemize}
Due to the high significance of $v$, OCSVM is also termed as $v-\text{SVM}$.

Since RSS algorithm can pinpoint the location of neighboring meters based on received electromagnetic signals, and OCSVM can detect anomaly in input data set, the combination of these two algorithms makes it possible to identify false nodes. In our security scheme, OCSVM utilizes packet size, frequency of data transmission and node identity along with the node position for meter authentication. The details have been provided in the next section.

\subsection{Entropy of a data packet}

Entropy is used to measure the uncertainty of a random variable or data packet. The more certain about a value, the smaller is the entropy  value.

The entropy for a sequence $S$

$H(s)=\sum_{S} P(S=x)\log_2 P(S=x)$

where $P(S=x)$ is the probability of taking $S$ a value over $x$.

If the size of a random variable or packet generated by the meter is $n$ bit, then the entropy and security strength of the data packet are $n$ and $2^n$ respectively.

\section{TRAFFIC FLOW PROCESS}

In this section, the privacy scheme implementation process and the data flow are described in detail.

In our architecture, we assume the following assumption:

\begin{itemize}
	
	\item  The Master and Auxiliary servers are independent and semi-trusted. However, the servers might physically be one but virtually divided into two servers.
	
	\item  The wireless communication links between servers and meters are not fully reliable.
	
	\item  The meters have limited memory and computational ability.
	
	\item  The control center has adequate computational ability.
	
	\item  The meters keep the records of position of neighboring meters, frequency of transmission, packet size and node identity. Frequency of transmission, node identity and packet size are extracted from the packet header. The node position is derived from electromagnetic signals using RSS based localization as explained in previous section IV(A).
	
	\item  Every meter transmits data at a constant transmission power.
	
	\item   The packet size is constant for every meter.
	
	\item   When a new meter is installed, it starts to record the position of the neighbor meters, frequency of data transmission, node identity  and packet size.
	
	\end{itemize}

\begin{figure}
\centering
  \includegraphics[width=0.5\textwidth,height=7cm]{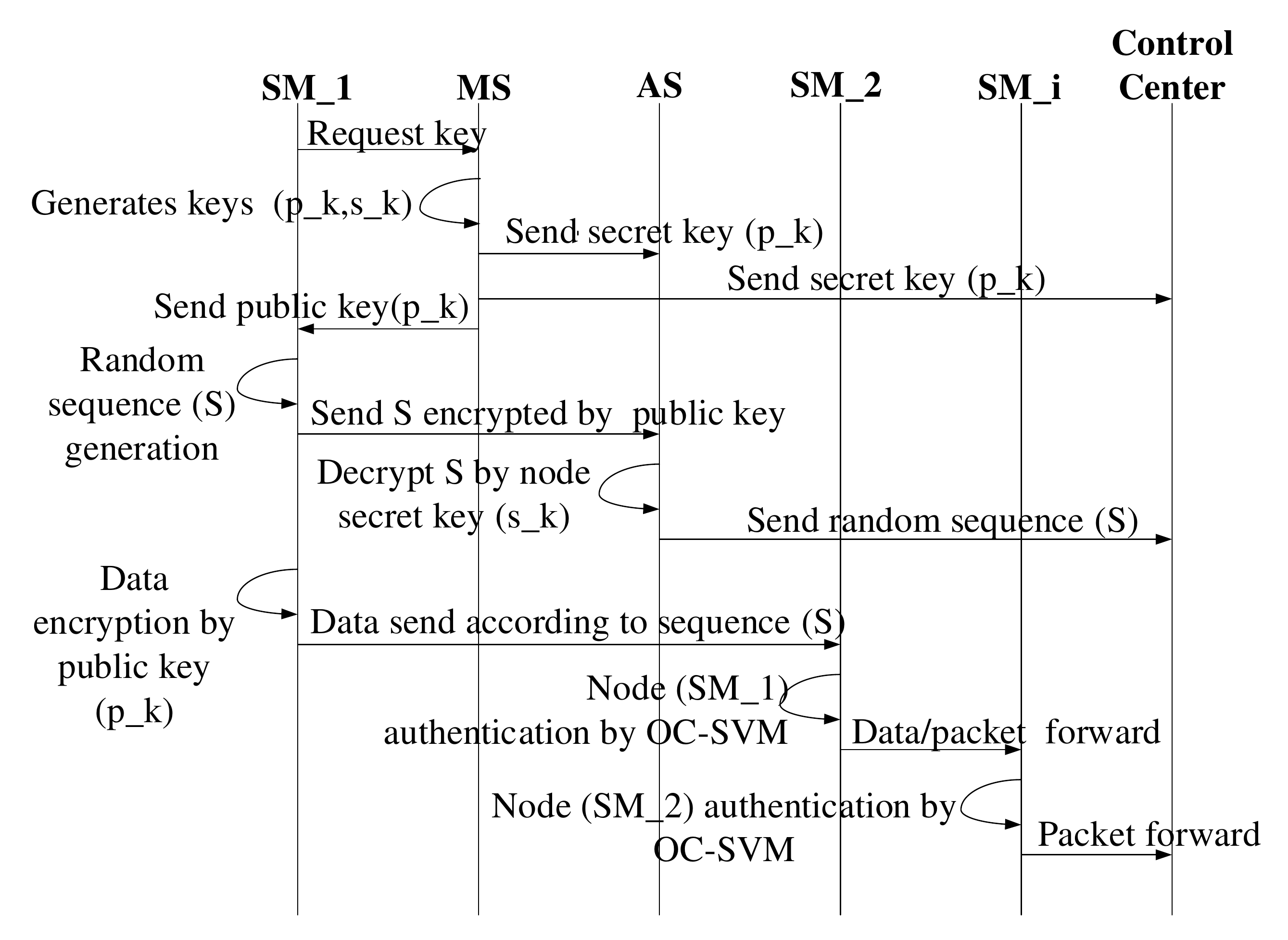}
  \caption{Data flow among various components of AMI.}
\end{figure}

\vspace{-0.4cm}
\subsection {Basic Notifications and definitions}

The notifications and definitions used in our algorithm are stated in Table I.\\

\begin{table}[h]
  
    \caption{Symbol notation.}
    \label{tab:table1}
    \begin{tabular}{p{4cm}p{4cm}}
    \toprule
    \multirow{1}{*}{\textbf{Notation}} & \textbf{Description} \\
      \midrule
      
      $SM_i$ & Smart meter $i$\\
      
      $A \mathcal{E}_i$ & Asymmetric encryption scheme for node/ meter $i$\\
      
      $k_i$ & Randomized key generation algorithm for node/ meter $i$\\
      
      $\mathcal{E}_i$ & Encryption algorithm for node $i$\\
       
      $D_i$& Deterministic decryption algorithm for node $i$\\
     
      $p_k$ & Public key\\
    
      $s_k$ & Secret key\\
     
      $M_i$ & Message/data of meter/node $i$  \\
      
      $C_i$ & Encrypted  data/ Message of node/ meter $i$\\
      
      $t$ & Time instance\\   
      
      $(m_1,m_2,.......,m_n)\in C$ & Segmented packets of message $C$\\
           
      $(s_1,s_2,....,s_n)\in S$ & Random sequences\\
                  
     $(p_1,p_2,......,p_n)\in P$ &  Probability of $ith$  packet transmission  at each time instant $t$\\   
 
      $\gamma$ & Path loss component \\   
 
      $\delta_l$ &  Variance of random noise \\ 
     
      PSO &  Particle swarm optimization \\ 
     
      One-class SVM &  One class Support Vector Machine\\   
     
      \bottomrule
    \end{tabular}
  
\end{table}

\vspace{-2\baselineskip}

\subsection {Data flow}

The data flow (as shown in Fig. 4) among the meters, servers and control center has be explained here in details.

\textbf {STEP 1: Initialization}

$SM_1$ sends a request for a public key before sending data. The Master server generates a public key and private key pair. The public key is unicasted to $SM_1$ for data encryption. Additionally, the corresponding private key is sent to the Auxiliary server and the control center.

Key generation by asymmetric algorithm:  

$A \mathcal{E}_1=(k_1,\mathcal{E}_1,D_1 )$

$k_1 \longrightarrow (p_{k1},s_{k1})$\\

\textbf {STEP 2: Encryption}

The $SM_1$ generates a random sequence ($S$). $S$ is encrypted by the received public key and is sent to the Auxiliary server. The Auxiliary server receives the cypher text and decodes it by the secret key. After that the Auxiliary server sends the sequence $S$ to the control center. At the same time, the consumption data is encrypted and processed by the following method:

Encryption of data: $p_{k1} \oplus M_1\longrightarrow C_1 $

Segmentation of the encrypted data into $n$ packets: $C_1 \longrightarrow (m_1, m_2, m_3......m_n)$

Ordering packets by random sequence: $(m_1, m_2, m_3......m_n)  \overset{S}{\longrightarrow} (h_1,h_2,h_3........h_n)$\\

\textbf {STEP 3: Data transmission}

The packets are transmitted by the  transmitting algorithm as explained in Algorithm 1:
\begin{algorithm}[h!]
\caption{Transmitting algorithm}\label{euclid}
\begin{algorithmic}[1]

\State \textbf {Initialization:}
\State Generate Random Sequence, $ S_t= (s_1,s_2,........,s_n) $ at time instant $t$

\If {$S_{t-1}==S_t$}
\State Go to Initialization
\Else
\State Proceed to next step
\EndIf
\State \textbf{end if}
\State Segment $C \longrightarrow (m_1,m_2,m_3.....m_n)$
\State Set $P_i= \frac{1}{S_i} \lhd P= (p_1,p_2,p_3,.......,p_n) $
where $p_1=\frac{1}{s_1},p_2=\frac{1}{s_2},p_3=\frac{1}{s_3}...........,p_n=\frac{1}{s_n} $
\State \textbf {Transmission:}
\State Transmit packet $m_i$ chosen from 
$(m_1,m_2,m_3,.....,m_n)$ with greater probability 
$\lhd$  probability distribution $(p_1,p_2,p_3,....,p_n)$ 
\State  \textbf {End}
\end{algorithmic}
\end{algorithm}

\textbf {STEP 4: Hop to hop data aggregation and forwarding}

Based on previous data receiving records containing node identity, sender node's position, frequency of data received and packet size- the node $SM_2$ verifies source node $SM_1$ and forwards data to the next node $SM_3$. OCSVM algorithm is used to authenticate the source node. The data aggregation and forwarding algorithm pseudocode is tabulated in Algorithm 2:
\begin{algorithm}[h!]
\caption{Data aggregation and forwarding}\label{euclid1}
\begin{algorithmic}[1]
\State \textbf {Initialization:}
\If {source meter $SM_i$ is new meter}
\State Record data transmission history (frequency of transmission, location of neighbor meter, packet size, node identity ) for time $h$
\State Send attach request to control center
\State Receive approval of new meter 
\Else 
\State Derive  $SM_i$'s location from received electromagnetic signal
\State Get frequency (time difference between two packet sending), node identity, packet size from header information of packet
\State Run OCSVM algorithm
\If {the new data and previous data belong to same group}
\State Forward data to the next meter
\Else
\State Cease data transmission and report to control center
\EndIf
\EndIf
\State \textbf{end if}
\State \textbf{end if}
\end{algorithmic}
\end{algorithm} 

%\vspace{1\baselineskip}

\textbf {STEP 5: Data retrieval}

The control center receives the randomized and encrypted packets and decodes them by the secret key $p_k$ and random sequence $S$.\\
Reordering the data: $(h_1,h_2,h_3,....,h_n)\overset{S}{\longrightarrow}(m_1,m_2,m_3.......,m_n)$
\\
Message unification:
 $ (m_1,m_2,m_3.......,m_n) \overset{}{\longrightarrow} C_1$
\\ 
Decryption: 
$ C_1\overset{p_k}{\longrightarrow} M_1 $

\section{Simulation result and Security Analysis}

\subsection{Simulation result}

To get insights into the localization of the meters, we used a suburban building block topology (Manhattan grid)\cite{j35} in which we used rectangle and hexagon shape as  area of interest (AOI). In an AOI, each edge is a known positioned meter and the center is the emitter (unknown positioned  meter). In Matlab simulation, we used $\gamma = 2.93 $ and $\sigma_l = 12 $ dB. Furthermore, for optimization PSO was used whereas residential pathloss model was considered for pathloss calculation.

%\begin{table} [h!]
%\caption{Simulation Parameter I}
%  \begin{tabular}{p{4cm}p{4cm}}
%    \toprule
%    \multirow{1}{*}{\textbf{Parameter}} & \textbf{Value} \\
%      \midrule
%       $\gamma$ & 2.93 \\   
%       $\sigma_l$ & 12 dB \\
%       Path loss model & Residential  \\ 
%       Optimization Technique & PSO \\ 
%    \bottomrule
%  \end{tabular}
%\end{table}

With an increase in the number of nodes (meters), the Mean Square Error (MSE) from the exact position of the meter decreases as illustrated in Fig. 5. Furthermore, as the variance of noise increases, the MSE also increases as depicted in Fig. 6. 

In the second part of simulation using Python, we use OCSVM for novelty/anomaly detection in a given data set. It takes node identity, node's position, frequency of transmission and packet size as new data set. The previous authenticate data set or fraction of new data set can be used as training data. In the simulation environment, we generate random 100 training observations, 20 normal regular observations and 20 abnormal observations. The red line (as shown in Fig. 7 and Fig. 8) is the learned decision function that separates abnormal data from the normal regular data. As shown in Fig.7, OCSVM draws a learned decision boundary based on  100 training clean data, of which 12 data falls outside the decision boundary erroneously. Following this, among the 20 regular novel data 3 data falls outside the boundary and all 20 abnormal data falls outside the decision boundary. So for 100 training data, OCSVM can detect anomaly with $100\%$ accuracy (0 error in novel abnormal detection). If we reduce the number of training samples to 20, the  performance  of OCSVM decreases (1 error in 20 novel abnormal detection) which is depicted in Fig. 8.

\begin{figure}[t!]
\centering
\includegraphics[width=1\linewidth]{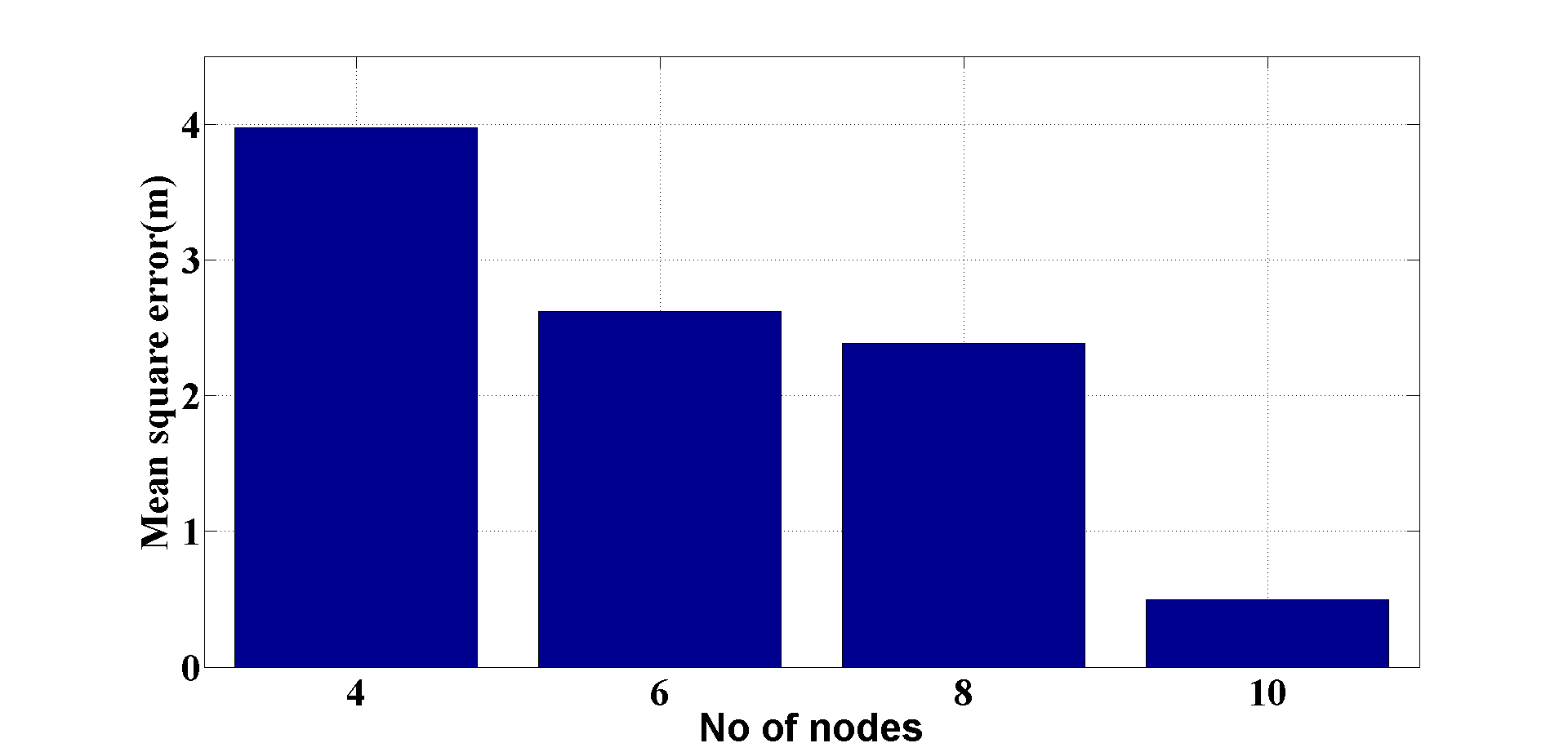}
\caption{MSE from the exact position of the meter vs. Number of nodes}\vspace{0 cm}
\label{fig:Packet}
\end{figure}

%\vspace{4\baselineskip}

\begin{figure}[t!]
\centering
\includegraphics[width=1\linewidth]{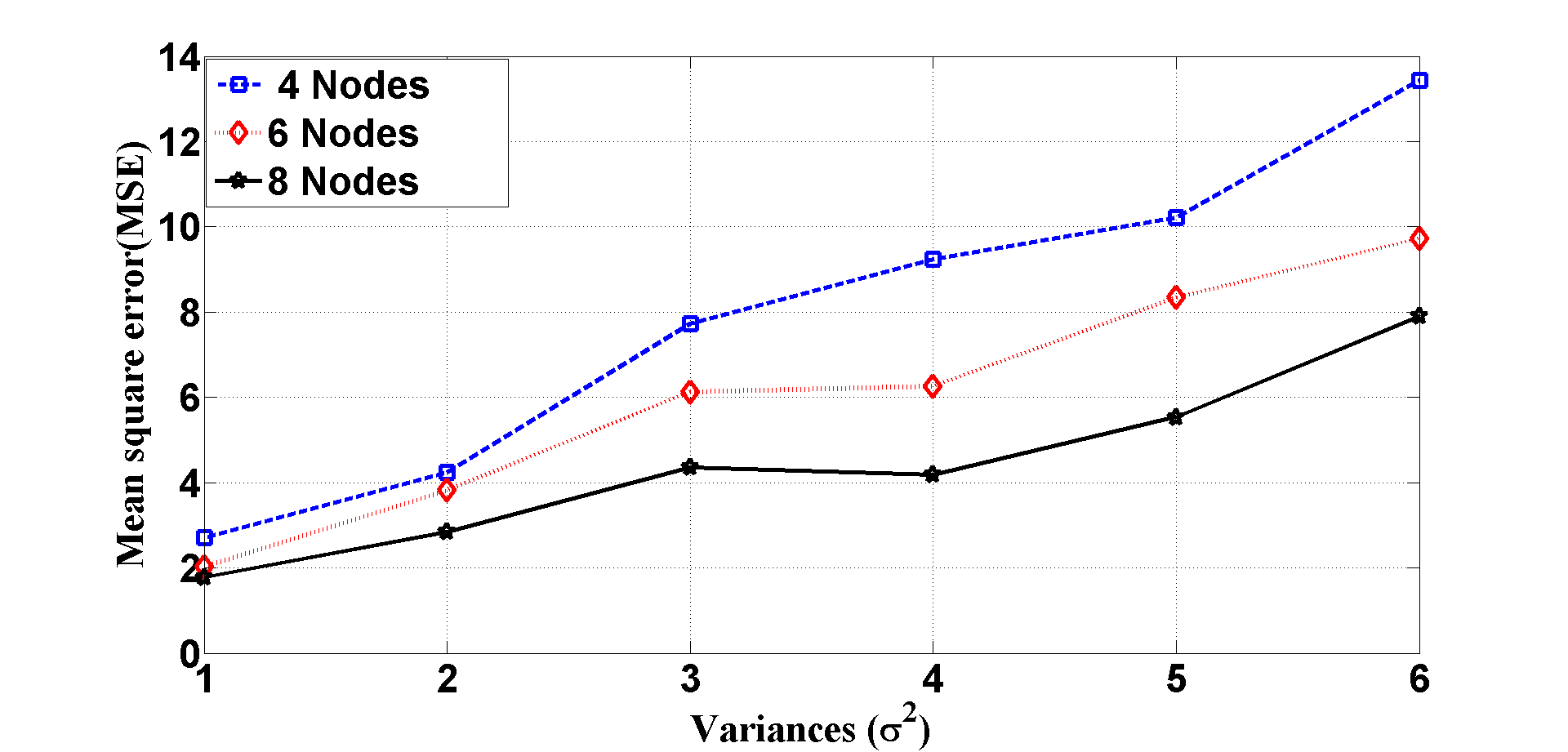}
\caption{ MSE from the exact position of the meter vs. Variance ($\sigma^2$) of Gaussian noise.}
\label{fig:Packet}
\end{figure}

\begin{figure*} [h!]
	\centering
	\subfigure[] { \label{fig:a}\includegraphics[width= 0.45\linewidth,height=4 cm]{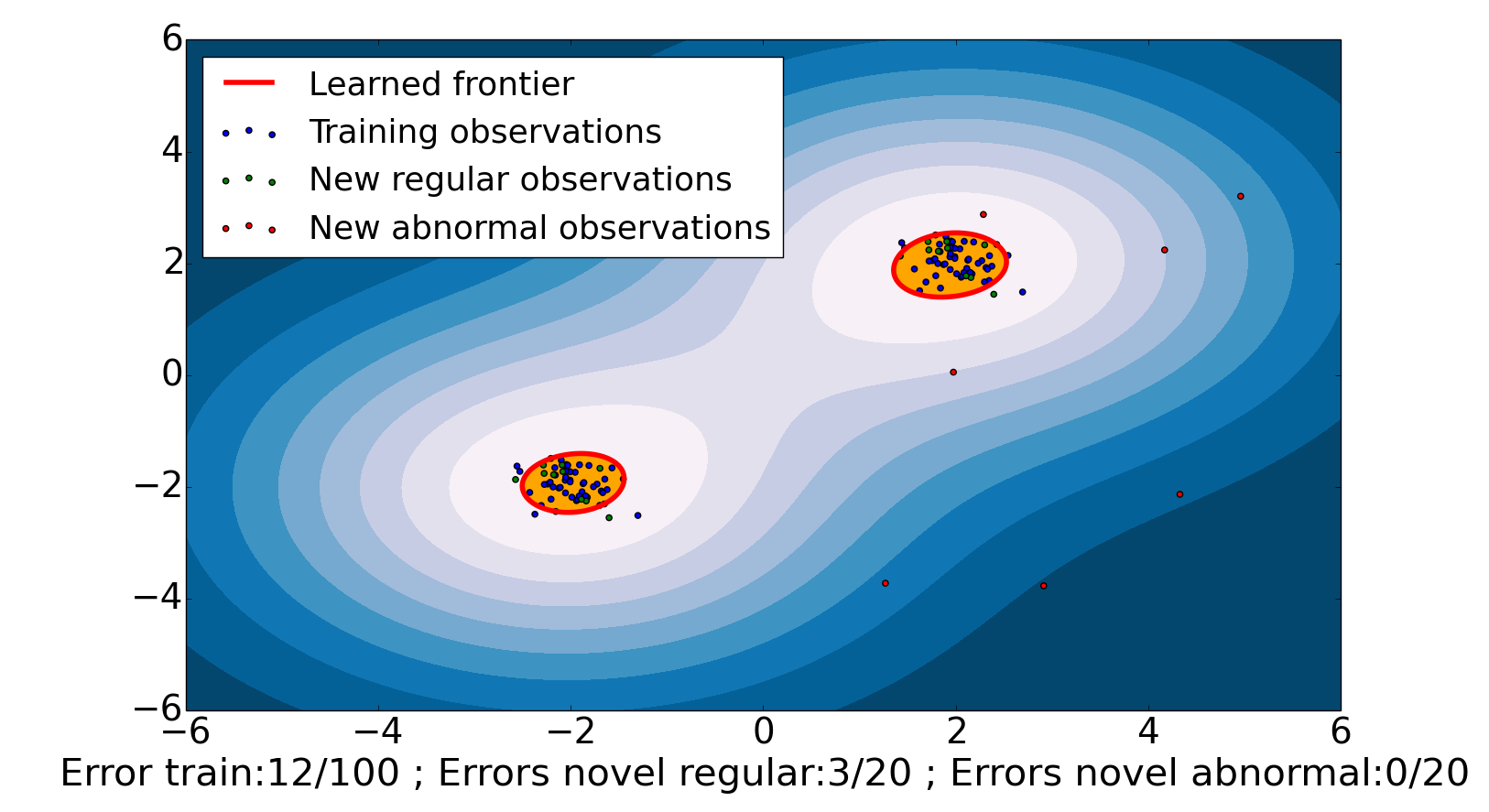}}\vspace{-0.15cm}
	\subfigure[] {\label{fig:b}\includegraphics[width= 0.45\linewidth,height=4 cm]{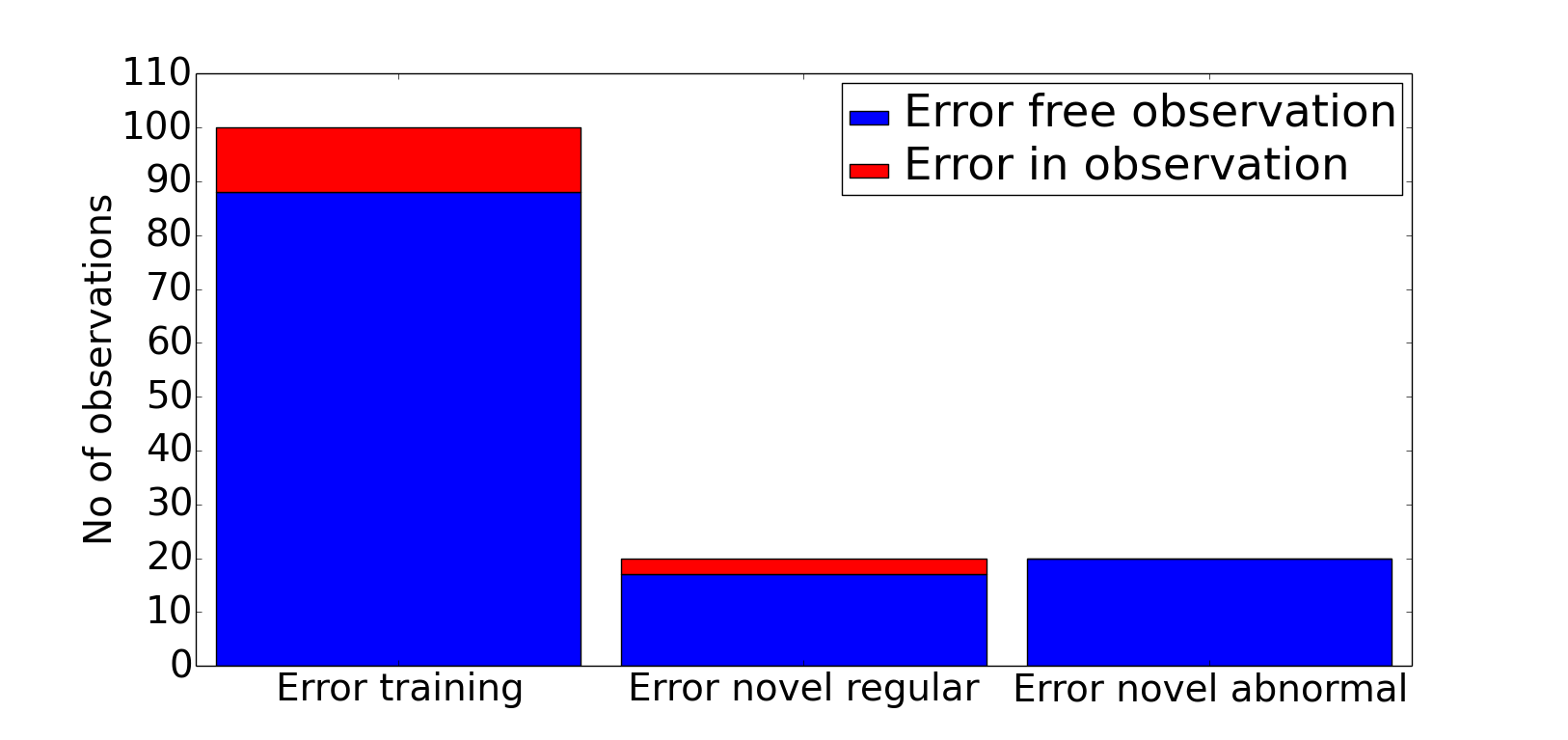}}\vspace{-0.15cm}
	\caption{Anomaly detection in 40 data with training data size 100. (a) Contour diagram of decision function. (b) Statistics of  error. }\vspace{0.05 cm}
	\label{fig:NE1}
\end{figure*}

\begin{figure*} [h!]
	\centering
	\subfigure[] { \label{fig:a}\includegraphics[width= 0.45\linewidth,height=4 cm]{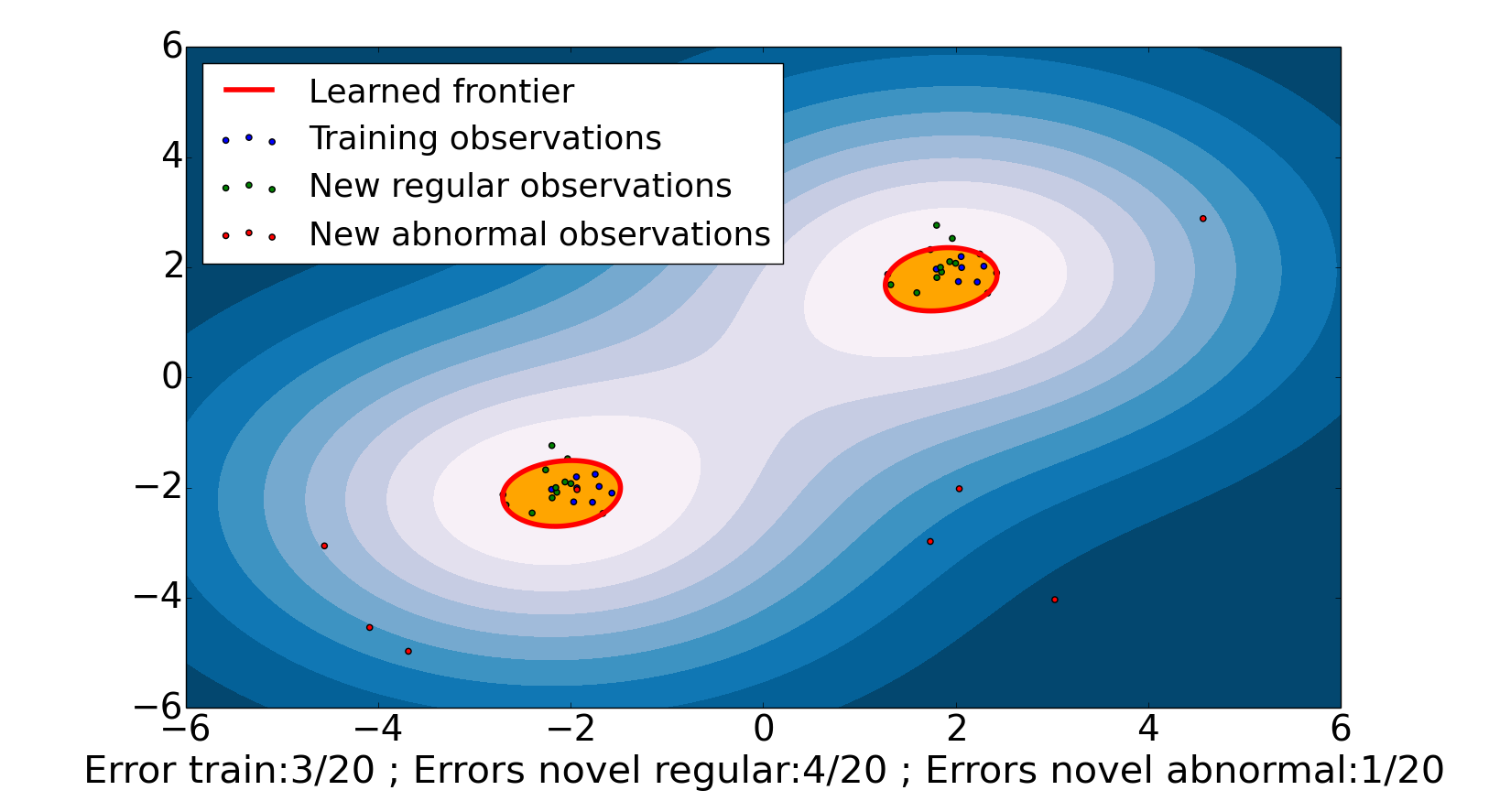}}\vspace{-0.15cm}
	\subfigure[] {\label{fig:b}\includegraphics[width= 0.45\linewidth,height=4 cm]{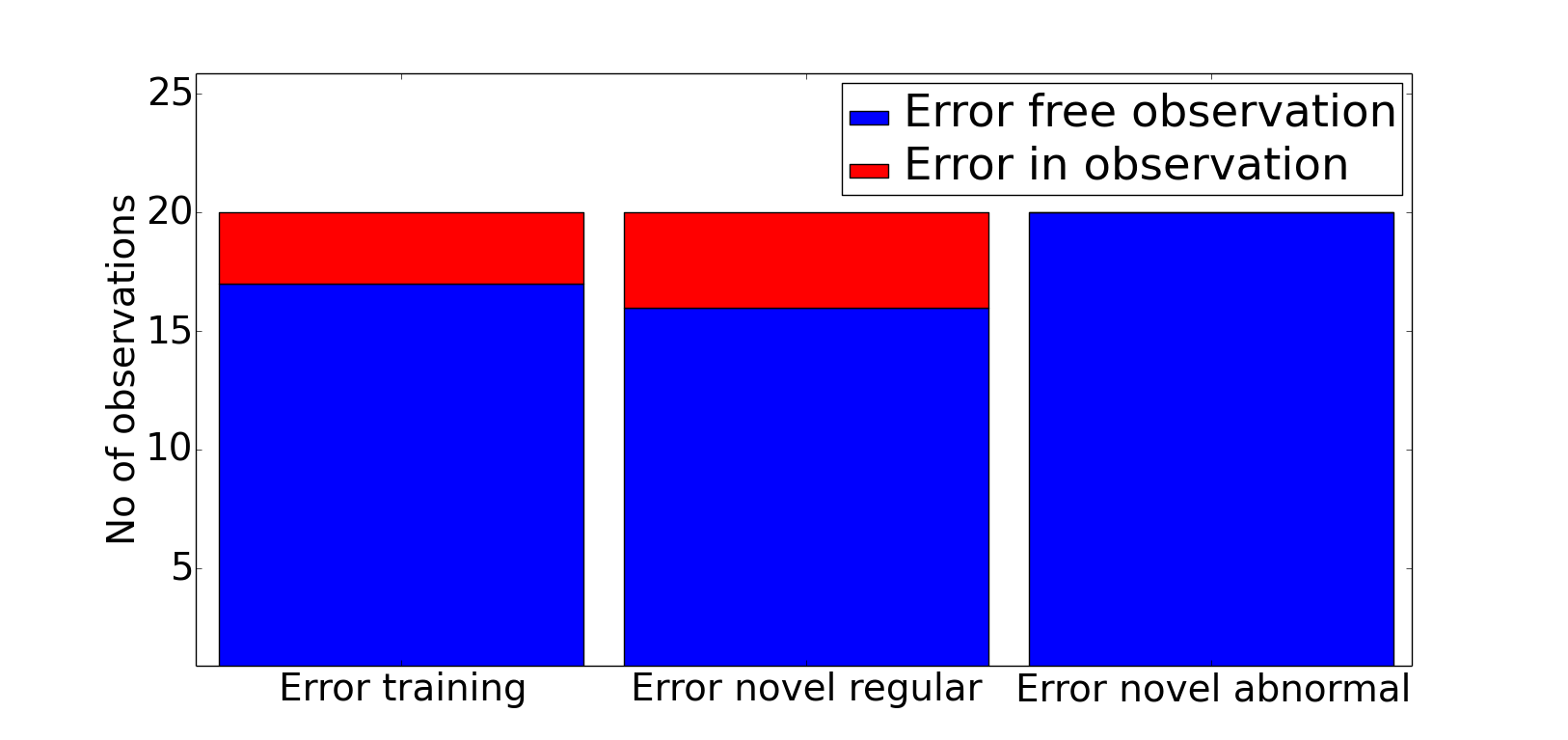}}\vspace{-0.15cm}
	\caption{ Anomaly detection in 40 data with training data size 20. (a) Contour diagram of decision function. (b) Statistics of  error. }\vspace{0.5cm}
	\label{fig:NE1}
\end{figure*}

%\begin{figure*}[h!]
%\centering
%\includegraphics[width= 0.7\linewidth,height=3.5 cm]{figure_int22.png}
%\caption{Anomaly detection in 40 data sample with training data size 100.}
%\label{fig: Packet11}
%\end{figure*}
%
%\begin{figure*}[h!]
%\centering
%\includegraphics[width= 0.7\linewidth, height=3.5 cm]{figure_int11.png}
%\caption{ Anomaly detection in 40 data sample with training data size 20.}
%\label{fig: Packet12}
%\end{figure*}

Now we use the polluted data for training purposes, which is a fraction of the input data. There is no separate clean training sample in this case. We also compare OCSVM with robust covariance estimator for different kinds of input data distribution. In this case,  we generate 100 samples, $10\%$ of which are outliers/abnormal. For a well-centered and elliptic input data set, OCSVM  and  robust covariance estimator show the same performance as illustrated in Fig. 9. For a bi-modal distribution of data, OCSVM shows better performance than the  robust covariance estimator. This is depicted in Fig. 10. For the non- Gaussian distributed input data, the performance of both the methods degrades, wherein OCSVM shows better performance than the  robust covariance estimator, illustrated in Fig. 11.

\subsection{Security strength analysis}

Let us assume, a meter sends a consumption unit packet of size 256 bit encrypted by 256 bit asymmetric key to the central database of control center. If the packet is subdivided into 32 blocks with each block having a size of 8, and a block is transmitted each time according to a random sequence, then the entropy is 256. The security strength of the data packet is $2^{256}$.

Furthermore, the security strength of a 256 bit asymmetric key is $2^{256/2}$. 

So, for 32 random sequenced packets and 256 bit asymmetric key,

Total security strength of the data packet = $2^{256} + 2^{256/2}$ 

Hence, a hacker needs a maximum ($2^{256} + 2^{256/2}$) number of iterations (tries) to decrypt a message, which is impractical.

\section{Conclusion}

In our security scheme, a two-level  security method has been proposed- data encryption and node authentication. In the data encryption level, encryption by asymmetric keys and randomization of data packets have been proposed. In the conventional key management system, only data encryption is used. On the other hand, in our scheme, randomization of packets along with data encryption ensures enhanced data security. Another contribution of our scheme is the introduction of node-to-node authentication by OCSVM, which utilizes  four variables- meter ID, frequency of data reception from a specific meter, packet size and meter position. The information of meter ID, data frequency and  packet size is easily extractable from the packet header. On the otherhand, for meter's position RSS is used and it is almost constant due to stationary position of the meters. 

For  TTP-to-smart meter communication, we need a bi-directional communication similar to meter data communication. Since  the communication between meters and servers  happens once per every session of  sending the meter data to the control center, it  doesn't hamper the normal traffic flow  between the meters and control center. Further, since a random sequence along with the asymmetric key is used to retrieve data in the control center, it also helps to verify data flow from a specific smart meter.

\begin{figure*}[h!]
\centering

\includegraphics[width= 0.6\linewidth,height=3.5 cm]{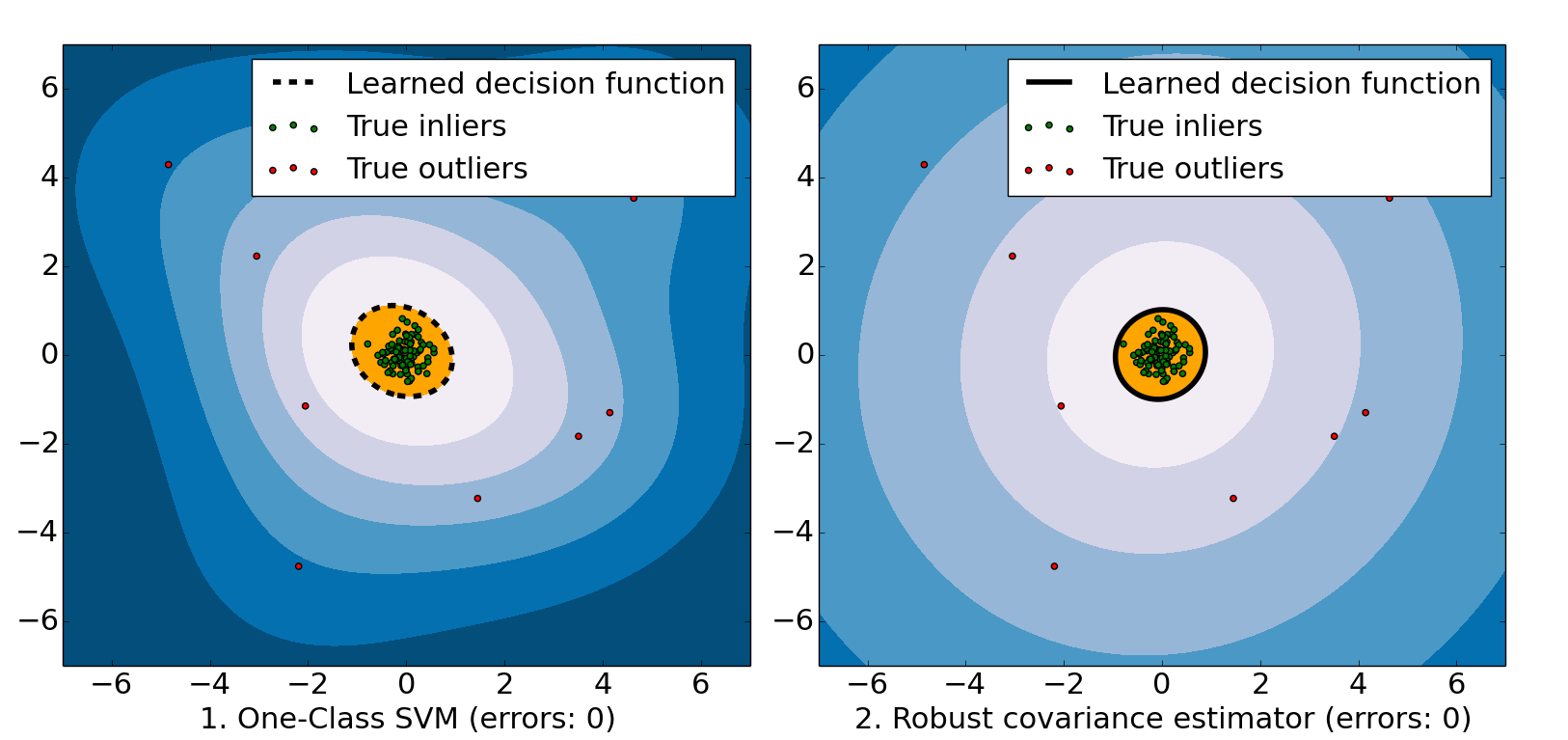}
\caption{ Comparison of OCSVM and robust covariance estimator for outliers (abnormal data) detection when inlier mode well-centered and elliptic.}\vspace{0.15cm}
\label{fig:Packet13}
\end{figure*}

\begin{figure*}[h!]
\centering
\includegraphics[width= 0.6\linewidth,height=3.5 cm]{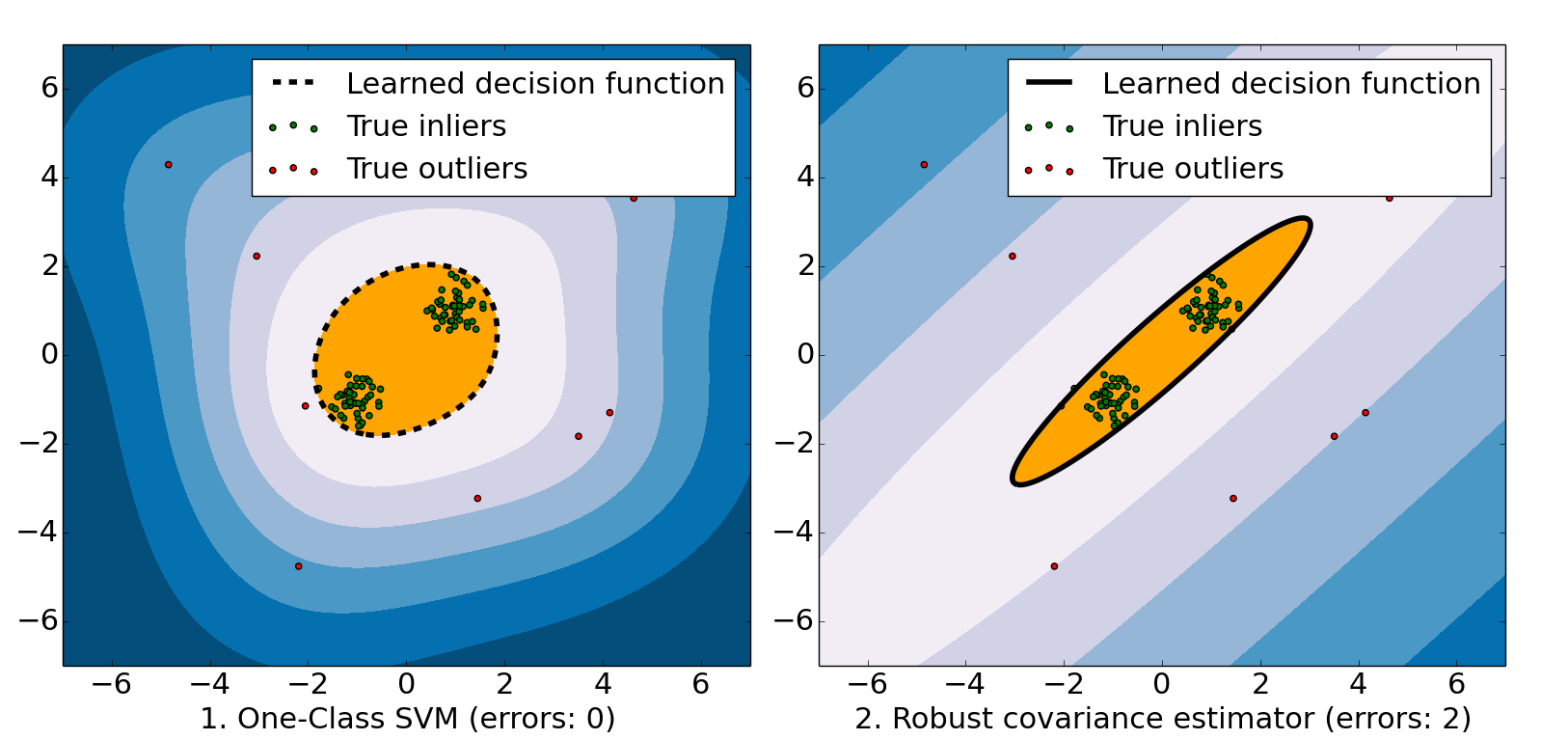}
\caption{ Comparison of OCSVM and robust covariance estimator for outliers (abnormal data) detection when inlier distribution is  bimodal.}\vspace{0.15cm}
\label{fig:Packet14}
\end{figure*}

\begin{figure*}[h!]
\centering
\includegraphics[width= 0.6\linewidth,height=3.5 cm]{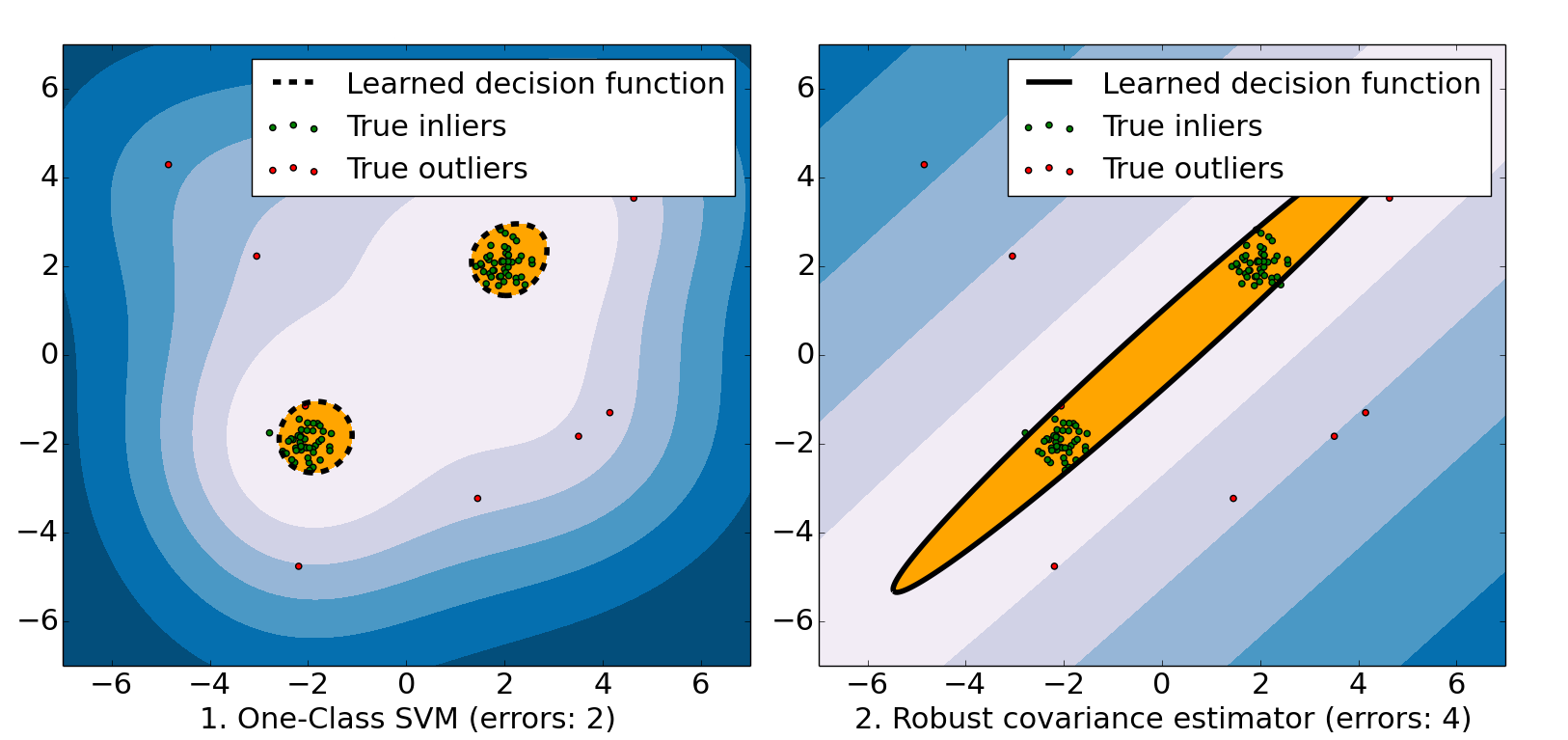}
\caption{ Comparison of OCSVM and robust covariance estimator for outliers (abnormal data) detection when inlier distribution is strongly non Gaussian.}
\label{fig:Packet15}
\end{figure*}

%\appendices
%\section{Proof of the First Zonklar Equation}
%Appendix one text goes here.
%
%% you can choose not to have a title for an appendix
%% if you want by leaving the argument blank
%\section{}
%Appendix two text goes here.
%
%
%% use section* for acknowledgement
%\section*{Acknowledgment}
%
%
%The authors would like to thank...

% Can use something like this to put references on a page
% by themselves when using endfloat and the captionsoff option.
\ifCLASSOPTIONcaptionsoff
  \newpage
\fi

\bibliographystyle{IEEEtran}
\bibliography{refkey}

% Generated by IEEEtran.bst, version: 1.13 (2008/09/30)
\begin{thebibliography}{10}
\providecommand{\url}[1]{#1}
\csname url@samestyle\endcsname
\providecommand{\newblock}{\relax}
\providecommand{\bibinfo}[2]{#2}
\providecommand{\BIBentrySTDinterwordspacing}{\spaceskip=0pt\relax}
\providecommand{\BIBentryALTinterwordstretchfactor}{4}
\providecommand{\BIBentryALTinterwordspacing}{\spaceskip=\fontdimen2\font plus
\BIBentryALTinterwordstretchfactor\fontdimen3\font minus
  \fontdimen4\font\relax}
\providecommand{\BIBforeignlanguage}[2]{{%
\expandafter\ifx\csname l@#1\endcsname\relax
\typeout{** WARNING: IEEEtran.bst: No hyphenation pattern has been}%
\typeout{** loaded for the language `#1'. Using the pattern for}%
\typeout{** the default language instead.}%
\else
\language=\csname l@#1\endcsname
\fi
#2}}
\providecommand{\BIBdecl}{\relax}
\BIBdecl

\bibitem{j28}
I.~Parvez, A.~Sundararajan, and A.~Sarwat, ``{Frequency band for HAN and NAN
  communication in Smart Grid},'' in \emph{Proc. on IEEE Sym. on Computational
  Intelligence Applications in Smart Grid (CIASG)}, Dec 2014, pp. 1--5.

\bibitem{j29}
F.~Granelli, D.~Domeniconi, N.~da~Fonseca, and B.~Tsetsgee, ``{On the Usage of
  WiFi and LTE for the Smart Grid},'' in \emph{Proc. Int. Conf. on Ubi-Media
  Computing and Workshops (UMEDIA)}, July 2014, pp. 1--5.

\bibitem{j30}
J.~Brown and J.~Khan, ``{Performance analysis of an LTE TDD based smart grid
  communications network for uplink biased traffic},'' in \emph{Proc. IEEE
  Globecom Workshops (GC Wkshps)}, Dec 2012, pp. 1502--1507.

\bibitem{j31}
L.~Li, H.~Xiaoguang, C.~Ke, and H.~Ketai, ``{The applications of WiFi-based
  Wireless Sensor Network in Internet of Things and Smart Grid},'' in
  \emph{Proc. IEEE Conf. on Industrial Electronics and Applications (ICIEA)},
  June 2011, pp. 789--793.

\bibitem{j1}
F.~Cleveland, ``{Cyber security issues for Advanced Metering Infrasttructure
  (AMI)},'' in \emph{Proc. IEEE conf. on Power and Energy Society General
  Meeting - Conversion and Delivery of Electrical Energy in the 21st Century},
  July 2008, pp. 1--5.

\bibitem{j17}
Z.~Lu, X.~Lu, W.~Wang, and C.~Wang, ``{Review and evaluation of security
  threats on the communication networks in the smart grid},'' in \emph{Proc.
  Conf. on MILITARY COMMUNICATIONS CONFERENCE (MILCOM)}, Oct 2010, pp.
  1830--1835.

\bibitem{j8}
Y.~Yan, Y.~Qian, and H.~Sharif, ``{A secure and reliable in-network
  collaborative communication scheme for advanced metering infrastructure in
  smart grid},'' in \emph{Proc. IEEE Int. conf. on Wireless Communications and
  Networking Conference (WCNC)}, March 2011, pp. 909--914.

\bibitem{j19}
S.~Das, Y.~Ohba, M.~Kanda, D.~Famolari, and S.~Das, ``{A key management
  framework for AMI networks in smart grid},'' \emph{IEEE Communications
  Magazine}, vol.~50, no.~8, pp. 30--37, August 2012.

\bibitem{j20}
Z.~Wan, G.~Wang, Y.~Yang, and S.~Shi, ``{SKM: Scalable Key Management for
  Advanced Metering Infrastructure in Smart Grids},'' \emph{IEEE Trans. on
  Industrial Electronics}, vol.~61, no.~12, pp. 7055--7066, Dec 2014.

\bibitem{j21}
------, ``{SKM: Scalable Key Management for Advanced Metering Infrastructure in
  Smart Grids},'' \emph{IEEE Trans. on Industrial Electronics}, vol.~61,
  no.~12, pp. 7055--7066, Dec 2014.

\bibitem{j22}
J.~Kamto, L.~Qian, J.~Fuller, and J.~Attia, ``{Light-weight key distribution
  and management for Advanced Metering Infrastructure},'' in \emph{Proc. IEEE
  Conf. on GLOBECOM Workshops (GC Wkshps)}, Dec 2011, pp. 1216--1220.

\bibitem{j12}
W.~Somkaew, S.~Thepphaeng, and C.~Pirak, ``{Data security implementation over
  ZigBee networks for AMI systems},'' in \emph{Proc. 11th Int. Conf. on
  Electrical Engineering/Electronics, Computer, Telecommunications and
  Information Technology (ECTI-CON}, May 2014, pp. 1--5.

\bibitem{j4}
Z.~Ismail, J.~Leneutre, D.~Bateman, and L.~Chen, ``{A Game Theoretical Analysis
  of Data Confidentiality Attacks on Smart-Grid AMI},'' \emph{IEEE J. on
  Selected Areas in Communications,}, vol.~32, no.~7, pp. 1486--1499, July
  2014.

\bibitem{j5}
S.~Amin, G.~Schwartz, A.~Cardenas, and S.~Sastry, ``{Game-Theoretic Models of
  Electricity Theft Detection in Smart Utility Networks: Providing New
  Capabilities with Advanced Metering Infrastructure},'' \emph{IEE J. on
  Control Systems}, vol.~35, no.~1, pp. 66--81, Feb 2015.

\bibitem{j34}
Y.~Yan, R.~Hu, S.~Das, H.~Sharif, and Y.~Qian, ``{An efficient security
  protocol for advanced metering infrastructure in smart grid},'' \emph{IEEE J.
  on Network}, vol.~27, no.~4, pp. 64--71, July 2013.

\bibitem{j55}
M.~Ali, E.~Al-Shaer, and Q.~Duan, ``{Randomizing AMI configuration for
  proactive defense in smart grid},'' in \emph{IEEE Int. Conf. on Smart Grid
  Communications (SmartGridComm)}, Oct 2013, pp. 618--623.

\bibitem{j13}
C.~Efthymiou and G.~Kalogridis, ``{Smart Grid Privacy via Anonymization of
  Smart Metering Data},'' in \emph{Proc. First IEEE Int. Conf. on Smart Grid
  Communications (SmartGridComm)}, Oct 2010, pp. 238--243.

\bibitem{j14}
N.~Yukun, T.~Xiaobin, C.~Shi, W.~haifeng, Y.~Kai, and B.~Zhiyong, ``{A security
  privacy protection scheme for data collection of smart meters based on
  homomorphic encryption},'' in \emph{Proc. IEEE conf. on EUROCON}, July 2013,
  pp. 1401--1405.

\bibitem{j32}
H.~So, S.~Kwok, E.~Lam, and K.-S. Lui, ``{Zero-Configuration Identity-Based
  Signcryption Scheme for Smart Grid},'' in \emph{Proc. First IEEE Int. Conf.
  on Smart Grid Communications (SmartGridComm)}, Oct 2010, pp. 321--326.

\bibitem{j33}
Y.~Yan, Y.~Qian, and H.~Sharif, ``{A secure and reliable in-network
  collaborative communication scheme for advanced metering infrastructure in
  smart grid},'' in \emph{Proc. IEEE Int. Conf. on Wireless Communications and
  Networking Conference (WCNC)}, March 2011, pp. 909--914.

\bibitem{j7}
R.~Bhatia and V.~Bodade, ``{Defining the framework for wireless-AMI security in
  smart grid},'' in \emph{Proc. Int. conf. on Green Computing Communication and
  Electrical Engineering (ICGCCEE)}, March 2014, pp. 1--5.

\bibitem{j15}
M.~Thomas, I.~Ali, and N.~Gupta, ``{A secure way of exchanging the secret keys
  in advanced metering infrastructure},'' in \emph{Proc. IEEE Int. Conf. on
  Power System Technology (POWERCON)}, Oct 2012, pp. 1--7.

\bibitem{j16}
P.-H. Hsu, W.~Tang, C.~Tsai, and B.-C. Cheng, ``{Two-Layer Security Scheme for
  AMI System in Taiwan},'' in \emph{Proc. Ninth IEEE Int. Symp. on Parallel and
  Distributed Processing with Applications Workshops (ISPAW)}, May 2011, pp.
  105--110.

\bibitem{j23}
N.~Patwari, J.~Ash, S.~Kyperountas, A.~Hero, R.~Moses, and N.~Correal,
  ``{Locating the nodes: cooperative localization in wireless sensor
  networks},'' \emph{IEEE Signal Processing Magazine}, vol.~22, no.~4, pp.
  54--69, July 2005.

\bibitem{j24}
L.~Xun and W.~Jianwen, ``{A cooperative framework for target tracking in
  Wireless sensor networks},'' in \emph{10th World Congress on Intelligent
  Control and Automation (WCICA)}, July 2012, pp. 4249--4254.

\bibitem{j25}
I.~Parvez, M.~Jamei, A.~Sundararajan, and A.~Sarwat, ``{RSS based loop-free
  compass routing protocol for data communication in advanced metering
  infrastructure (AMI) of Smart Grid},'' in \emph{Proc. on IEEE Sym. on
  Computational Intelligence Applications in Smart Grid (CIASG)}, Dec 2014, pp.
  1--6.

\bibitem{j26}
L.~Clemmensen, T.~Hastie, and B.~Ersboll, ``{Sparse Discriminat Analysis},''
  \emph{Technometrics}, vol.~53, no.~4, pp. 406--413, June 2008.

\bibitem{j27}
\BIBentryALTinterwordspacing
I.~Czogiel, K.~Luebke, M.~Zentgraf, and C.~Weihs,
  ``\BIBforeignlanguage{English}{Localized linear discriminant analysis},'' in
  \emph{\BIBforeignlanguage{English}{{Advances in Data Analysis}}}, ser.
  Studies in Classification, Data Analysis, and Knowledge Organization,
  R.~Decker and H.-J. Lenz, Eds.\hskip 1em plus 0.5em minus 0.4em\relax
  Springer Berlin Heidelberg, 2007, pp. 133--140. [Online]. Available:
  \url{http://dx.doi.org/10.1007/978-3-540-70981-7_16}
\BIBentrySTDinterwordspacing

\bibitem{j35}
J.~Markkula and J.~Haapola, ``{LTE and hybrid sensor-LTE network performances
  in smart grid demand response scenarios},'' in \emph{Proc. IEEE Conf. on
  Smart Grid Communications (SmartGridComm)}, Oct 2013, pp. 187--192.

\end{thebibliography}


% Generated by IEEEtran.bst, version: 1.13 (2008/09/30)
\begin{thebibliography}{10}
\providecommand{\url}[1]{#1}
\csname url@samestyle\endcsname
\providecommand{\newblock}{\relax}
\providecommand{\bibinfo}[2]{#2}
\providecommand{\BIBentrySTDinterwordspacing}{\spaceskip=0pt\relax}
\providecommand{\BIBentryALTinterwordstretchfactor}{4}
\providecommand{\BIBentryALTinterwordspacing}{\spaceskip=\fontdimen2\font plus
\BIBentryALTinterwordstretchfactor\fontdimen3\font minus
  \fontdimen4\font\relax}
\providecommand{\BIBforeignlanguage}[2]{{%
\expandafter\ifx\csname l@#1\endcsname\relax
\typeout{** WARNING: IEEEtran.bst: No hyphenation pattern has been}%
\typeout{** loaded for the language `#1'. Using the pattern for}%
\typeout{** the default language instead.}%
\else
\language=\csname l@#1\endcsname
\fi
#2}}
\providecommand{\BIBdecl}{\relax}
\BIBdecl

\bibitem{j42}
V.~Gungor, D.~Sahin, T.~Kocak, S.~Ergut, C.~Buccella, C.~Cecati, and G.~Hancke,
  ``{Smart Grid Technologies: Communication Technologies and Standards},''
  \emph{Industrial Informatics, IEEE Transactions on}, vol.~7, no.~4, pp.
  529--539, Nov 2011.

\bibitem{j1}
F.~Cleveland, ``{Cyber security issues for Advanced Metering Infrasttructure
  (AMI)},'' in \emph{Proc. IEEE conf. on Power and Energy Society General
  Meeting - Conversion and Delivery of Electrical Energy in the 21st Century},
  July 2008, pp. 1--5.

\bibitem{j17}
Z.~Lu, X.~Lu, W.~Wang, and C.~Wang, ``{Review and evaluation of security
  threats on the communication networks in the smart grid},'' in \emph{Proc.
  Conf. on MILITARY COMMUNICATIONS CONFERENCE (MILCOM)}, Oct 2010, pp.
  1830--1835.

\bibitem{j8}
Y.~Yan, Y.~Qian, and H.~Sharif, ``{A secure and reliable in-network
  collaborative communication scheme for advanced metering infrastructure in
  smart grid},'' in \emph{Proc. IEEE Int. conf. on Wireless Communications and
  Networking Conference (WCNC)}, March 2011, pp. 909--914.

\bibitem{j19}
S.~Das, Y.~Ohba, M.~Kanda, D.~Famolari, and S.~Das, ``{A key management
  framework for AMI networks in smart grid},'' \emph{IEEE Communications
  Magazine}, vol.~50, no.~8, pp. 30--37, August 2012.

\bibitem{j21}
Z.~Wan, G.~Wang, Y.~Yang, and S.~Shi, ``{SKM: Scalable Key Management for
  Advanced Metering Infrastructure in Smart Grids},'' \emph{IEEE Trans. on
  Industrial Electronics}, vol.~61, no.~12, pp. 7055--7066, Dec 2014.

\bibitem{j41}
I.~Parvez, A.~Islam, and F.~Kaleem, ``{A key management-based two-level
  encryption method for AMI},'' in \emph{PES General Meeting | Conference
  Exposition, 2014 IEEE}, July 2014, pp. 1--5.

\bibitem{j12}
W.~Somkaew, S.~Thepphaeng, and C.~Pirak, ``{Data security implementation over
  ZigBee networks for AMI systems},'' in \emph{Proc. 11th Int. Conf. on
  Electrical Engineering/Electronics, Computer, Telecommunications and
  Information Technology (ECTI-CON}, May 2014, pp. 1--5.

\bibitem{j4}
Z.~Ismail, J.~Leneutre, D.~Bateman, and L.~Chen, ``{A Game Theoretical Analysis
  of Data Confidentiality Attacks on Smart-Grid AMI},'' \emph{IEEE J. on
  Selected Areas in Communications,}, vol.~32, no.~7, pp. 1486--1499, July
  2014.

\bibitem{j5}
S.~Amin, G.~Schwartz, A.~Cardenas, and S.~Sastry, ``{Game-Theoretic Models of
  Electricity Theft Detection in Smart Utility Networks: Providing New
  Capabilities with Advanced Metering Infrastructure},'' \emph{IEE J. on
  Control Systems}, vol.~35, no.~1, pp. 66--81, Feb 2015.

\bibitem{j34}
Y.~Yan, R.~Hu, S.~Das, H.~Sharif, and Y.~Qian, ``{An efficient security
  protocol for advanced metering infrastructure in smart grid},'' \emph{IEEE J.
  on Network}, vol.~27, no.~4, pp. 64--71, July 2013.

\bibitem{j55}
M.~Ali, E.~Al-Shaer, and Q.~Duan, ``{Randomizing AMI configuration for
  proactive defense in smart grid},'' in \emph{IEEE Int. Conf. on Smart Grid
  Communications (SmartGridComm)}, Oct 2013, pp. 618--623.

\bibitem{j13}
C.~Efthymiou and G.~Kalogridis, ``{Smart Grid Privacy via Anonymization of
  Smart Metering Data},'' in \emph{Proc. First IEEE Int. Conf. on Smart Grid
  Communications (SmartGridComm)}, Oct 2010, pp. 238--243.

\bibitem{j14}
N.~Yukun, T.~Xiaobin, C.~Shi, W.~haifeng, Y.~Kai, and B.~Zhiyong, ``{A security
  privacy protection scheme for data collection of smart meters based on
  homomorphic encryption},'' in \emph{Proc. IEEE conf. on EUROCON}, July 2013,
  pp. 1401--1405.

\bibitem{j32}
H.~So, S.~Kwok, E.~Lam, and K.-S. Lui, ``{Zero-Configuration Identity-Based
  Signcryption Scheme for Smart Grid},'' in \emph{Proc. First IEEE Int. Conf.
  on Smart Grid Communications (SmartGridComm)}, Oct 2010, pp. 321--326.

\bibitem{j33}
Y.~Yan, Y.~Qian, and H.~Sharif, ``{A secure and reliable in-network
  collaborative communication scheme for advanced metering infrastructure in
  smart grid},'' in \emph{Proc. IEEE Int. Conf. on Wireless Communications and
  Networking Conference (WCNC)}, March 2011, pp. 909--914.

\bibitem{j7}
R.~Bhatia and V.~Bodade, ``{Defining the framework for wireless-AMI security in
  smart grid},'' in \emph{Proc. Int. conf. on Green Computing Communication and
  Electrical Engineering (ICGCCEE)}, March 2014, pp. 1--5.

\bibitem{j15}
M.~Thomas, I.~Ali, and N.~Gupta, ``{A secure way of exchanging the secret keys
  in advanced metering infrastructure},'' in \emph{Proc. IEEE Int. Conf. on
  Power System Technology (POWERCON)}, Oct 2012, pp. 1--7.

\bibitem{j16}
P.-H. Hsu, W.~Tang, C.~Tsai, and B.-C. Cheng, ``{Two-Layer Security Scheme for
  AMI System in Taiwan},'' in \emph{Proc. Ninth IEEE Int. Symp. on Parallel and
  Distributed Processing with Applications Workshops (ISPAW)}, May 2011, pp.
  105--110.

\bibitem{j36}
V.~N. Vapnik, \emph{{Statistical learning theory}}, 1st~ed.\hskip 1em plus
  0.5em minus 0.4em\relax Wiley, Sep. 1998.

\bibitem{j37}
R.~Zhang, S.~Zhang, S.~Muthuraman, and J.~Jiang, ``{One Class Support Vector
  Machine for Anomaly Detection in the Communication Network Performance
  Data},'' in \emph{Proceedings of the 5th Conference on Applied
  Electromagnetics, Wireless and Optical Communications}, ser.
  ELECTROSCIENCE'07, 2007, pp. 31--37.

\bibitem{j35}
J.~Markkula and J.~Haapola, ``{LTE and hybrid sensor-LTE network performances
  in smart grid demand response scenarios},'' in \emph{Proc. IEEE Conf. on
  Smart Grid Communications (SmartGridComm)}, Oct 2013, pp. 187--192.

\end{thebibliography}

\end{document}